\documentclass[aps,twocolumn,prl,amsmath,amssymb,floatfix,superscriptaddress]{revtex4}

\usepackage{graphicx}
\usepackage{dcolumn}
\begin{document}

\bibliographystyle{revtex}


\title{Theoretical proposal predicting anomalous magnetoresistance and quadratic Hall effect in the partially ordered state of MnSi}


\author{B. Binz}
\email{binzb@berkeley.edu} \affiliation{Department of Physics, University of California,  Berkeley, CA 94720, USA}
\author{A. Vishwanath}\affiliation{Department of Physics, University of California,  Berkeley, CA 94720, USA}


\date{\today}

\newcommand{\eref}[1]{(\ref{#1})}

\newcommand{\bv}[1]{{\bf #1}}
\newcommand{\uv}[1]{{\bf \hat #1}}

\newcommand{\be}{\begin{equation}}
\newcommand{\ee}{\end{equation}}

\begin{abstract} In [B. Binz, A. Vishwanath and V. Aji, {\it Phys. Rev. Lett.} {\bf 96}, 207202 (2006)], a magnetic structure that breaks time reversal symmetry in the {\em  absence} of  net magnetization was proposed as an explanation for the high pressure "partially ordered" state of MnSi. Here we make explicit the anomalous magneto-transport properties of such a state: a  magnetoresistivity which is linear and a Hall conductance which is quadratic in the applied magnetic field. Field cooling procedures for obtaining single domain samples are  discussed. The anomalous effects are elaborated in the case of three geometries chosen to produce experimentally unambiguous signals of this unusual magnetic state; e.g., it is predicted that a field in z-direction induces an anisotropy in the x-y plane. Another geometry leads to a Hall voltage {\em parallel} to the magnetic field.
 \end{abstract}

\maketitle

{\em Introduction:}
Recently, we proposed novel magnetic states, called bcc helical spin crystals  \cite{binz06}, which may occur  in certain non-centrosymmetric compounds, e.g. in the "partial-order" phase of MnSi \cite{pfleiderer04}. These bcc spin crystals break time reversal symmetry globally, but nevertheless have a vanishing overall magnetization. 
 Based on their symmetries, we predict that  bcc spin crystals lead to distinctive anomalous signatures in magnetotransport and the Hall effect.

{\em Normal magnetoresistance and Hall effect:}
In a normal paramagnetic metal with cubic symmetry, the resistivity tensor $\rho$, defined as $E_a=\rho_{ab}\, j_b$,
behaves in the presence of a weak applied magnetic field $\bv B$  as:
\be
\rho_{ab}=(\rho_0+\alpha' |\bv B|^2)\delta_{ab}+\alpha B_aB_b-\rho_H\epsilon_{abc}B_c,
\ee
where $\rho_0$ is the isotropic zero-field resistivity, $\rho_H$ is responsible for the ordinary Hall effect and $\alpha,\alpha'$ determine the longitudinal and transverse magnetoresistivity. This form is imposed by the behavior under spatial rotations and time reversal. 
The tensor $\rho$ 
 splits into  symmetric and antisymmetric parts, $\rho=\rho^s+\rho^a$.
The symmetric part 
 is the magnetoresistivity, which depends on $\rho_0, \alpha$ and $\alpha'$. It is diagonal in an appropriate basis. The antisymmetric part gives rise to the Hall field $E^H_a=1/2(\rho_{ab}-\rho_{ba}) j_b$, or equivalently
\be
\bv E^H=\rho_H\,\bv B\times \bv j.\label{hall0}
\ee
Onsager relations imply that  the magnetoresistivity part is even under time-reversal (i.e. even in $\bv B$) and the Hall-effect part is odd under time-reversal (odd in $\bv B$).

{\em Bcc spin crystals:}
bcc spin crystals \cite{binz06} are characterized by the global time-reversal symmetry breaking order parameter  $S=\langle M_xM_yM_z\rangle\neq0$, where $M_a$ (for $a=x,y,z$) are components of the local magnetization along the cubic crystal axes and $\langle\ldots\rangle$ means averaging over the sample. More precisely, they satisfy
\begin{eqnarray}
\langle M_a\rangle&=&0,\\
\langle M_a M_b \rangle&\propto& \delta_{ab},\\
\langle M_a M_bM_c \rangle&=& S\, |\epsilon_{abc}|.\label{S}
\end{eqnarray}
In the absence of magnetic field, any bcc spin crystal state is doubly degenerate, as there is a state with $S>0$ and a degenerate time-reversal symmetry partner with $S<0$ (equal magnitude, opposite sign).
The energy splitting between the two time-reversal symmetry partners in a magnetic field is
$
\Delta E 
\propto S\,B_x B_y B_z,
$
 e.g. a field along $[111]$ ($[\bar1 \bar1 \bar1 ]$) favors $S<0$ ($S>0$) or vice versa.

{\em Obtaining a Single Domain Sample:}
The experimental characteristics 
 described below 
 depend crucially on having a single domain of the time reversal symmetry broken states ($S>0$) or ($S<0$). If the sample is cooled without magnetic field, we expect domains with $S>0$ and $S<0$ to be created in equal proportions such that the anomalous effects 
 cancel out in any macroscopic measurement. 
Given the above splitting between the ground states in a field, one way of obtaining a single domain sample would be to cool in a magnetic field directed along the  $[111]$ axis, and then turn the field to zero inside the partially ordered state.

{\em Anomalous magnetoresistivity and Hall effect:}
 bcc spin crystal systems do {\em not} exhibit a zero-field Hall effect. Nevertheless, their magneto-transport is anomalous  due to 
 Eq.~\eref{S}. For example the magnetoresistivity 
 obtains
an unusual component {\em linear} in the field:
\be
\Delta \rho^s_{ab}=\beta\, \langle M_aM_bM_c\rangle B_c.\label{beta}
\ee
Although $\Delta\rho^s$  is even under time-reversal as required by the Onsager relations, it is {\it linear} in $\bv B$. Similarly, the Hall conductance has a piece that is {\em quadratic} in the field:
\be
\Delta\rho^a_{ab}=-\frac12 \gamma\, \epsilon_{abc} \langle M_cM_dM_e\rangle B_dB_e.\label{gamma}
\ee
  Equation \eref{hall0} now becomes
\be
\bv E^H=\rho_H\,\bv B^{\rm eff}\times \bv j, \label{EHall}
\ee
 where
\be
\bv B^{\rm eff}=\bv B + \frac{\gamma}{\rho_H}\, S \left(\,B_y B_z\,,\,B_z B_x\,,\, B_x B_y\,\right).
\label{Beff}
\ee
 
In the following, we discuss three examples of experimental arrangements that may be particularly suitable to measure the anomalous  contributions of Eqs.~\eref{beta} and \eref{gamma}.

\begin{figure}[t]
\includegraphics[scale=0.5]{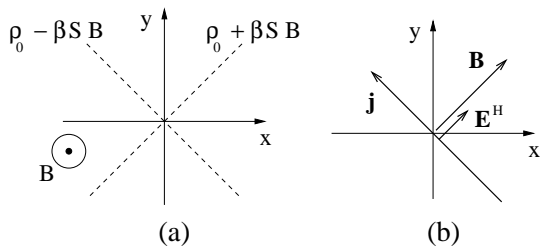}
\caption{(a) Principal axes of magnetoresistivity for a magnetic field $\bv B \parallel \uv z$. (b) If a field is applied along  $[110]$ and a current is run along $[1\bar10]$, there is an anomalous component of the Hall field in the direction of $\bv B$. The ordinary component $B\rho_H\uv z$ is not shown.} 
\label{combination}
\end{figure}

{\em Field along [001]:} The field along $\uv z$ induces an anisotropic magnetoresistivity in the $x-y$ plane. The principal axes of resistivity are $[110]$ and $[1\bar10]$ respectively, with eigenvalues $\rho_0 \pm \beta S B+\alpha' B^2$  [Fig.~\ref{combination} (a)].  Thus the field-dependent {\em anisotropy} is the anomalous signature sought for in this measurement.

 {\em Field along  [110]:} Consider the case $B_z=0$, $B_x=B_y=B/\sqrt{2}$ with a current running along the $[1\bar10]$ direction. 
 The $[1\bar10]$ direction is a principal axis of magnetoresistivity with eigenvalue $\rho_0+\alpha'B^2$, i.e. there are no anomalous contributions to magnetoresistivity.
 But due to Eqs.~\eref{EHall} and \eref{Beff},
\be
\bv E^H=B(\rho_H\uv z-\gamma S \bv B) j.
\ee
Thus, $\bv E^H$ requires an anomalous component in field-direction [Fig.~\ref{combination} (b)].
 A Hall voltage {\em parallel} to the applied magnetic field is the predicted anomalous signature  for this geometry.

\begin{figure}[t]
\includegraphics[scale=0.5]{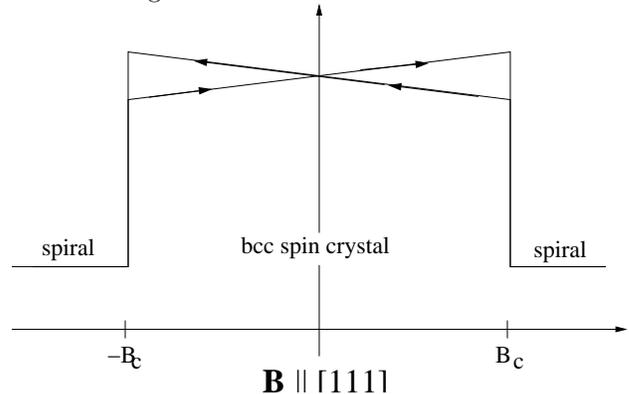}
\caption{Schematic sketch of the bow-tie hysteresis expected for both the magnetoresistivity and the Hall effect in the case of a magnetic field along the  $[111]$ direction. It is assumed that at  critical fields $\pm B_c$, the system makes a transition into a different magnetic state, e.g. the helical spin-density wave state \cite{nakanishi80}.
}
\label{bowtie}
\end{figure}

{\em Field along  [111]:} In the case $\bv B\parallel [111]$, 
the magnetoresistivity  in field-direction is 
\be
\rho_{\parallel}=\rho_0+\frac2{\sqrt{3}}\beta S B+(\alpha+\alpha')B^2,\label{para}
\ee
 where $B/\sqrt{3}=B_x=B_y=B_z$,   and the magnetoresistivity in any direction perpendicular to the field is
\be
\rho_{\perp}=\rho_0-\frac1{\sqrt{3}}\beta S B+\alpha'B^2.\label{perp}
\ee
The Hall effect becomes
\be
\rho_H\, \bv B^{\rm eff}=(\rho_H+\frac1{\sqrt{3}}\gamma S B)\,\bv B.\label{hall}
\ee
 Since the field along $[111]$ splits the degeneracy of the two time-reversal symmetry partners, it may eventually switch the sign of  $S$. This would result in a bow-tie shaped hysteresis as shown in Fig.~\ref{bowtie}, both in the magnetoresistivity [Eqs.~\eref{para}, \eref{perp}] and in the amplitude of the Hall effect  [Eq.~\eref{hall}]. 
 Thus, the field-induced hysteresis in both magnetoresistance  and Hall effect  is the anomalous signature sought for in this experiment.

{\em Conclusion:}
Observation  of  any of these anomalous transport properties would be an unambiguous experimental proof of  macroscopic time-reversal symmetry breaking and,
 together with the observed neutron scattering peaks along $\langle 110\rangle$ directions \cite{pfleiderer04}, it  would identify the ``partial order'' state of MnSi as a bcc spin crystal 
 beyond doubt.

\end{document}